\documentstyle[twocolumn,aps]{revtex}
\begin{document}
\baselineskip = 1\baselineskip  
\title{\Large \bf About Charge Density Wave \\
for Electromagnetic Field-Drive}
\author{Beno\^\i t T. Guay\cite{btg1}}
\address{Qu\'{e}bec, November 24, 1999}
\maketitle

\centerline{\bf Abstract}
\begin{abstract}
To generate a propulsive force without propellant and external couplings,
it has been shown that two confined macroscopic and time-varying charge
density waves well separated in space are needed. Here, some physical 
conditions will be proposed to support and maintain these particular 
collective modes of charge distributions.
\end{abstract}

\section{Introduction}
Within the framework of classical electrodynamics, it has been shown \cite{btg}
how an electromagnetic propulsive force and, in particular, an electric 
(conservative) propulsive force can be generated without propellent
mass and external couplings by using two confined, time-varying, neutral
and macroscopic charge density waves (CDW). These CDW own a same symmetry
axis, are adequately separated in space and have a relative temporal 
phase-shift. This last one controls the propulsive force's intensity. 

From far fields point of view, these CDW are able to induce an asymmetry
into the space distribution of the far fields momentum variation rate along
the symmetry
axis. They can do that because the relative temporal phase-shift controls
the space distribution of constructive and destructive interferences of
far fields produced by the two CDW \cite{btg}. So, this relative temporal 
phase-shift controls the asymmetry. When this last one is created, an 
electromagnetic propulsive force along the symmetry axis is generated and
applied on both CDW in a same direction. Such propulsive effect is 
impossible in statics because fields'
interferences can be produced only with time-varying fields. Because this
propulsive force is generated by a spatial asymmetry in the (electromagnetic)
field, it is a propulsion driven by the electromagnetic field or more
simply an electromagnetic field-drive (EFD). 

In our first paper \cite{btg} we have used the CDW concept in a theoretical
way. Actually, nothing has been said about
the material or the conductive fluid needed to sustain a neutral macroscopic
charge
density wave. The only thing we have mentioned was this CDW is a longitudinal
(i.e. $\phi$ direction in cylindrical coordinates) charge oscillation mode,
it has a wave number ``n", it
oscillates at frequency $\omega$ and it is pinned
(circular standing wave) inside a ring made with an electrical conductor.
In this simplified model,
we have used two identical planar filiform rings with radii R', placed in
vacuum
and separated by a distance D along the z axis. Planes of rings were 
perpendicular to the z axis; the symmetry axis, the thrust axis.
In a more
realistic way, rings have a cross section $R_o$ 
smaller than D and R' according to section 4 in \cite{btg}. However, we have
never mentioned that a relation must exist (dispersion relation) between
n and $\omega$ and what is this relation. Furthermore, what are needed 
conditions to support and maintain
a time-varying CDW able to create the desired propulsive effect?
Is it possible to use solid rings? Metallic ones? Or what else? 
In this work, we would like 
to give preliminary and partial answers to some of those
questions. 

\section{A Longitudinal Plasma Mode}
A time-varying longitudinal CDW involves a time-varying longitudinal charge
separation among opposite charges. In that case, there must be a restoring
force among these charges and consequently, this creates a collective 
oscillation mode (i.e. longitudinal plasma mode) at plasma
frequency ${\omega}_p$ \cite{jac1,jor1,lor1,ash1,kit1}. So, to sustain a
large amplitude of charge separations
in a neutral conductor or, more generally, in a conductive ``fluid" and then
support and maintain sources of large electric fields, our frequency $\omega$
must be close to (at least equal or greater than) the ``resonant" frequency
${\omega}_p$. Thus, we will get an
appropriate CDW (n$\neq$0) if each neutral conductive fluid of our two rings
is a neutral plasma.

The other reason to use a $\omega$ $>$ ${\omega}_p$ is this. In a sense
${\omega}_p$ can be considered as a cut-off \cite{jor2}.
So, if $\omega$ is greater than this cut-off, fields created by one conductive fluid in a given
ring will penetrate deeply inside the conductive fluid of the other ring
to create propulsive effect throughout 
the ring's cross section for non-filiform rings (i.e. torus for instance).
Actually, if $\omega$ $<$ ${\omega}_p$ fields generated by one
ring will remain near the surface of the other; they will be mostly reflected
by this one and they will be nearly zero inside of it except to its surface.
In such a case, the thrust's amplitude will be
limited and restricted to the rings' surface. In addition, this will increase
the probability of cold emission like in a metal (see below) because fields
must be relatively strong (i.e. at least about 100kv) to get a good thrust
\cite{btg}. So, things like that can reduce the propulsive effect. 

According to the model in \cite{btg}, the value of $\omega$ must be in the
range of radio frequency or TV range. Consequently, our plasma must have
a ${\omega}_p$ in these ranges too. However, if n$=$0 there are no
charge separations at all; we have only a uniform longitudinal current on
each ring. In this last case, we don't need a longitudinal plasma mode; a
neutral conductive fluid with a ${\omega}_p$ much larger
than $\omega$ can be used. But let's remember this, if n$=$0 the propulsive
force has no electric contribution (i.e. no conservative part), only a
magnetic one (i.e. a dissipative part because radiative) and we know that
this last contribution
has a poor efficiency according to section 6 in \cite{btg}. 

For our purpose and for now, at least four limits or conditions 
must be considered in a neutral
plasma. The first is related to the collision rate $\it{f}$. Our macroscopic
time-varying CDW is a collective oscillation mode; a longitudinal plasma
mode. Collisions break ``coherence" among charges' motions and then break
the collective oscillation and, the
CDW itself can be destroyed. To get collective oscillations, we must have
$\it{f}$ $\ll$ ${\omega}_p$ $\sim$ $(e^2n_o/m_e{\epsilon}_o)^{1/2}$ (SI units)
\cite{jac1,jor1,lor1,ash1,kit1}; $m_e$ is the effective electron's mass,
e the electron's charge, $n_o$ the electron density when n$=$0 
(for electronic plasma with heavy positive ions as uniform background) 
and ${\epsilon}_o$ is the vacuum
electrical permittivity. We have to mention that $\it{f}$ increases with
$n_o$ and temperature (see below).

The second limit is associated with the wave number or the wave length 
of the charge density in a
conductive fluid. For us, this is related to ``n". There is an upper limit for
this wave number. Above this limit, the CDW cannot oscillate; the damping
(i.e. Landau damping \cite{jac2}) is too strong and thus,
CDW does not exist (it's too ``viscous"). In a nondegenerate conductive fluid,
like an ionized gas with relatively small density of electrons and ions for
instance, this upper wave number is the Debye
wave number $k_D$ given by ${k_D}^{-1}$ $\sim$ $(T_e/n_o)^{1/2}$ cm 
\cite{jac3,jor3} ($n_o$ in cm$^{-3}$). $T_e$ is the electrons' 
temperature (in Kelvin); a measure of their mean kinetic energy. 
An electron within ${k_D}^{-1}$ cannot move easily (``viscous" area) but
outside, it can. So, if the wave
length of our CDW is larger than ${k_D}^{-1}$, the damping won't exist or
it will be weak or quite weak and then, this CDW will survive and will be
able to oscillate.

In a degenerate neutral conductive fluid like an electron gas in a solid
metal at low temperature (i.e. low compared to the Fermi energy $E_F$ 
\cite{ash2,kit2}), $k_D$ is replaced by the Fermi wave number $k_F$ 
\cite{jac3}. In that case, the typical kinetic energy is $E_F$ not $T_e$. 

In our situation, we need a neutral
plasma with ${\omega}_p$ $\sim$ 100 MHz (radio frequency as order of 
magnitude) and a ${k_{DorF}}^{-1}$ smaller than about ${10}^{-2}$ cm. 
${10}^{-2}$ cm is
a lower limit for the wave length of our CDW; a macroscopic length scale for
which our classical approach in \cite{btg} is certainly correct. With solid
alkali metals like Li, Na, etc. or solid noble metals
like Cu, Ag, Au, ${k_F}^{-1}$ respects the above condition. For example,
solid copper (Cu) at room
temperature ($\sim$ 300K), ${k_F}^{-1}$ $\sim$ ${10}^{-8}$ cm 
\cite{kit3,ash3}. But the problem with solid metals like alkali (or noble)
is their ${\omega}_p$ belong to ultraviolet frequency range 
($\sim$ ${10}^{15}$Hz) \cite{ash4,kit4}. The reason for such a big value
is a large $n_o$ ($\sim$ ${10}^{22}$ cm$^{-3}$) \cite{kit4} and a very
small effective mass of charge carrier (i.e. electron). So, solid
metals can bee used only if n $=$ 0 (i.e. uniform
currents on rings) according to above discussion ($\omega$ $<$ $\omega_p$).

For instance, if n $=$ 0, we could use two metallic and 
solid torus (i.e. planar rings with cross section $R_o$ 
smaller than D and R' according to above), fixed apart with
distance D by some adequat isolators and placed in good vacuum at ``room
temperature". However,
one possible problem with metals is the cold emission \cite{yav1}; when
fields applied over metallic crystal become relatively strong, electrons
(carriers) can be expelled outside the crystal by ``quantum 
tunneling". In that case,
the charge's momentum of
carriers won't be given to the whole crystal along the thrust axis so,
the momentum transfer
efficiency will be diminished and then, the propulsion too. Furthermore,
with metals we
will have $\omega$ $<$ ${\omega}_p$ and, as mentioned above, this is
limitative. 

With n $\neq$ 0, we need something else. For instance an ionized gas; 
a neutral conductive gas formed by
electrons and ions with a smaller electron density: $n_o$ $\sim$ ${10}^{8}$
cm$^{-3}$. In that case this neutral plasma has a
${\omega}_p$ in the range that we want according to its expression given
above. On the other hand, we want
a relatively ``cold" plasma because we wish to satisfy the condition 
$\it{f}$ $\ll$ ${\omega}_p$ and also because we
want to avoid any complications about plasma confinement (``walls"). For
example, let's consider a
temperature $T_e$ between 1000K to 10000K. Such values for $T_e$ and $n_o$
give us a ${k_D}^{-1}$ $\sim$ ${10}^{-3}$ to ${10}^{-2}$ cm according to
the above expression so, they 
give us a classical
plasma (i.e. nondegenerate electrons gas where classical statistics can be
applied) quite similar to the
ionosphere's one \cite{psr}. Actually, at 90km into ionosphere, collision rate
is $\it{f}$ $\sim$ ${10}^6$Hz and at 300km, $\it{f}$ $\sim$ ${10}^3$Hz 
\cite{jor4}. Such last values respect the preceding inequality between
$\it{f}$ and ${\omega}_p$. This doesn't
mean ion species we need must be the same as the ionosphere's ones. 
Best ion species we need is
another issue. But it shows that such a kind of plasma exist. So, a priori, a neutral
ionized gas with relatively ``low"temperature, ${10}^3$ to ${10}^4$K, and
low electrons density, $n_o$ $\sim$ ${10}^8$ cm$^{-3}$, (i.e. a cold plasma)
could be a good candidate for our purpose when n $\neq$ 0. 

Let's take an example to get an order
of magnitude of the propulsive force when a cold plasma gas is under 
consideration. Let us consider a lithium
gas with
electrons density $n_o$ $\sim$ ${10}^8$ cm$^{-3}$ and electrons temperature
$T_e$ $\sim$ $5000$K. According to above expressions,
${\omega}_p$ $\sim$ $564$MHz and ${k_D}^{-1}$ $\sim$
$7$$\times$${10}^{-3}$ cm.
By simplicity, let's imagine 
all atoms of lithium are ionized such as Li $\rightarrow$ Li$^+$ $+$
e$^-$. Atomic weight of Li is about 6.9a.m.u. so lithium mass density is
about: $n_o$$\times$6.9$\times$1.66$\times$${10}^{-27}$kg $\sim$
${10}^{-18}$kg/cm$^3$. (Of course, this doesn't take into account the mass
of confining ``walls"). Mass of Li$^+$ is about ${10}^4$ times larger than
the one of $e^-$. So, ion Li$^+$ is at rest compared to e$^-$; only 
electrons move at frequency $\omega$ along $\phi$ direction. Now to get
an order of magnitude of
the propulsive force, we can use the Coulomb force
expression. Coulomb force
is one of main contributions (conservative part) to the thrust in \cite{btg}. 
So, in these conditions if we consider a small volume of
1cm$^3$ of charges on each ring (or torus), the force we can get 
between these small volumes if D $=$ 0.1m (same order of magnitude than
the one used in \cite{btg})
is given approximately by (1cm$^3$)$^2$$\cdot$$\bigl($ ${n_o}^{2}$${e}^{2}$/$4{\pi}$${\epsilon}_o$$D^2$ $\bigr)$ $\sim$ 
${10}^{-10}$N. This evaluation is a $\it{maximum}$
one because it doesn't take into account destructive 
interferences among fields produced by positive and negative charges in a
same CDW and applied over charges in the other CDW. 

The reason for such a small force is the relatively small value of $n_o$.
If we increase $n_o$, condition ${k_D}^{-1}$ $\ll$ ${10}^{-2}$ cm will be
always satisfied but certainly not ${\omega}_p$ $\sim$ 100MHz. However, if
we use an ``ionic plasma" instead of an ``electronic one" as in the above
example, we will have ${\omega}_p$ $\sim$ 
$({q}^2n_o/m_i{\epsilon}_o)^{1/2}$ 
and ${k_D}^{-1}$ $\sim$ $(T_i/n_o)^{1/2}$ cm where $n_o$ is now the ions
density, $T_i$ the ions temperature, $m_i$ is the reduced mass of ions 
and $q$, their charge. Consequently,
if $n_o$ is increased, we will keep ${\omega}_p$ fixed if we take an 
appropriate reduced mass $m_i$ larger than $m_e$. Let's give an example. 

Let's take Li $+$ Cl $\rightarrow$ Li$^+$ $+$ Cl$^-$. Ion 
chlorine Cl$^-$ is about 5 times heavier than ion Li$^+$ so, $m_i$ 
$\sim$ $m_{Li}$ $=$ $11.4$$\times$${10}^{-27}$kg, $q$ = $e$ and 
$T_i$ $\sim$ $T_{Li}$. As before we take same temperature $T_{Li}$ $\sim$ 5000K. Now
to get the same plasma frequency; ${\omega}_p$ $\sim$ 564MHz, we must
take $n_o$ $\sim$ $1.3$$\times$${10}^{12}$cm$^{-3}$. In that case, 
${k_D}^{-1}$ $\sim$ $6.2$$\times$${10}^{-5}$cm and 
(1cm$^3$)$^2$$\cdot$$\bigl($${n_o}^{2}$${e}^{2}$/$4{\pi}$${\epsilon}_o$$D^2$$\bigl)$ $\sim$ 
$3.9$$\times$${10}^{-2}$N with the same D as before. However, 
condition $\it{f}$ $\ll$ ${\omega}_p$
is not respected. We can evaluate $\it{f}$ by using its expression 
\cite{reif,reichl} for
an ideal gas (i.e. low density and pressure). One has $\it{f}$ $\sim$ 
$n_o$$\bar v$${\sigma}_{Cl}$
$=$ $n_o$$({8k_BT_{Li}}/{\pi m_{Li}})^{1/2}$${\sigma}_{Cl}$ $\sim$ 6.2GHz. 
$k_B$ is the Boltzmann's constant, ${\sigma}_{Cl}$ 
$\sim$ $\pi$(${k_D}^{-1}$)$^2$ is the scattering
cross section of the screened chlorine ion and $\bar v$ is the mean speed of lithium ion;
this velocity is close to the relative velocity between lithium and chlorine
ions. Finally, mass
density is $n_o$($6.9$$+$$35.4$)$\times$$1.66$$\times$
${10}^{-27}$kg $\sim$ $9.1$$\times$${10}^{-14}$kg/cm$^3$.
So, as we can see, the choice of ion species is quite important.

The neutral plasma gas must be ionized by some external source 
(at the beginning at least) but,
because temperature is relatively small, after a specific time there are
recombinations among
electrons and ions (or ions-ions) and then a radiation (named secondary here)
is emitted. The primary radiation is the one emitted by the longitudinal
plasma oscillations of both CDW at frequency $\omega$ $> \atop \sim$ 
$\omega_p$. Other kinds of secondary
radiations can also be emitted like breaking radiation (bremsstrahlung)
\cite{yav2,jac5} and spectral radiation
coming from excited atoms (not ionized). Recombinations among charges imply
that a third limit has to be considered in our neutral plasma. This limit
is given by ${\it{f}}_r$ $\ll$ ${\omega}_p$ where ${\it{f}}_r$ is the
recombination rate between negative and positive charges. Clearly, this
quantity depends on
electrons (or ions) density $n_o$ and electrons (or ions) temperature $T_e$
(or $T_i$). ${\it{f}}_r$ increases when temperature decreases because kinetic
energy of opposite charges (i.e. their thermal energy) becomes smaller than
their potential energy (i.e. mutual attraction). This is why
temperature, on the other hand, cannot be too small.  

\section{Anisotropic Conductive Gas}
According to the model given in \cite{btg}, charges must be well confined
along the z direction (i.e. the thrust direction) and along the $\rho$ 
direction in some restricted regions (i.e. ``filiform" rings). 
So, some constraints have to exist to
maintain charges in these limited areas along those directions. These
constraints have to ensure also the momentum transfer from charges to confining
``walls", specially along z. In that sense, the conductive
fluid (or gas) must be strongly
anisotropic; charges can move easily along $\phi$ but should be nearly 
``at rest" along z and $\rho$ directions.

Now, to get an appropriate anisotropic conductive gas (ionic and cold
plasma gas), the cross section's radii $R_o$ of a ring (or torus) 
must be smaller or equal to ${k_D}^{-1}$ so, the fourth limit is 
$R_o$ ${< \atop \sim}$ ${k_D}^{-1}$ $\sim$ $6.2$ $\times$ ${10}^{-5}$ cm
(using preceding value of chlorine-lithium gas) so, a ``micro-torus" with a
relatively large radii R'. 
The reason is this. Any charges inside ${k_D}^{-1}$,
around the heavier ion; the Cl$^-$ in our previous example, are 
in ``viscous" area. This is true for Li$^+$ ions and for
induced dipoles of the dielectric ``walls" (see below). Consequently, with
the above limit, any
relative motions between Cl$^-$ and Li$^+$ along z and $\rho$ are quite well
limited and this is true also among Cl$^-$ and dipoles, induced by this ion,
inside the internal surface of the dielectric walls along those
directions. 

In addition, the wall of this micro-torus must be a good dielectric.
The neutral ionized gas will fill the micro-torus. The
dielectric wall must be transparent to primary and secondary
radiations. This is obvious for primary fields according to above; fields
must reach the gas. But 
it is also important for the secondary to maintain a fixed temperature and
get and sustain an equilibrium between ionization and recombination. 
Furthermore, this dielectric wall
must be able to support high mechanical stress and relatively high 
temperature. 

\section{Conclusion}
In this paper, a well confined neutral ionized gas at relatively low 
density and
temperature (i.e. a nondegenerated conductive gas; a ``cold plasma") 
is proposed as a substrate in which a CDW
(n $\neq$ 0) can be sustained; the CDW needed to produce
a conservative propulsive force, according
to the model given in our first work.

Up to now, cold plasma is probably the most appropriate material able to
create conservative propulsive force and meet conditions
given in this paper. But, plasma stability, plasma confinement, momentum
transfer from accelerated charges to the confining ``walls" along the thrust
axis, choice of best ion species and dispersion relation are certainly 
complicated issues to deal with in the near-term. In addition, the fourth
condition is difficult to satisfy from a technological point of view now.
On the other hand, as shown in \cite{btg}, this model (i.e. rings and
the specific charge and current density distributions used; the CDW) has
a poor efficiency. For all of
these reasons, modifications to this model (i.e. to charge distributions) 
are needed to get a more efficient
and realistic near-term EFD.

\end{document}